\documentclass{article}

     \PassOptionsToPackage{numbers, compress}{natbib}
     \usepackage[final]{neurips_2020_ml4ps}

\usepackage[utf8]{inputenc} % allow utf-8 input
\usepackage[T1]{fontenc}    % use 8-bit T1 fonts
\usepackage{hyperref}       % hyperlinks
\usepackage{url}            % simple URL typesetting
\usepackage{booktabs}       % professional-quality tables
\usepackage{amsfonts}       % blackboard math symbols
\usepackage{nicefrac}       % compact symbols for 1/2, etc.
\usepackage{microtype}      % microtypography
\usepackage{amsmath}
\usepackage{graphicx}
\usepackage{color}
\usepackage{soul}
\usepackage[ruled,vlined]{algorithm2e}

\DeclareMathOperator*{\argmin}{arg\,min}

\title{Meta Variational Monte Carlo
}

\author{
Tianchen Zhao$^{1}$, \,James Stokes$^{2}$, \,Oliver Knitter$^{1}$, \,Brian Chen$^{1}$, \,Shravan Veerapaneni$^{1,2}$ \vspace{0.05in}\\ 
  $^{1}$Department of Mathematics, University of Michigan, Ann Arbor, MI 48109\vspace{0.05in} \\
  $^{2}$Flatiron Institute, Simons Foundation, New York, NY 10010 \vspace{0.05in}\\
  \texttt{\{ericolon, knitter, chenbri, shravan\}@umich.edu}\\
  \texttt{jstokes@flatironinstitute.org}\\
}

\begin{document}

\maketitle

\begin{abstract}
An identification is found between meta-learning and the problem of determining the ground state of a randomly generated Hamiltonian drawn from a known ensemble. A model-agnostic meta-learning approach is proposed to solve the associated learning problem and a preliminary experimental study of random Max-Cut problems indicates that the resulting Meta Variational Monte Carlo accelerates training and improves convergence.
\end{abstract}

\section{Introduction}
This paper concerns the development of heuristic approximation algorithms for determining the ground state (minimal eigenpair) of certain large random Hermitian matrices drawn from known ensembles of relevance to Physics. Concretely, we focus on the real vector subspace of Hermitian matrices consisting of 2-local, $n$-qubit  Hamiltonia, which are enumerated by a number of parameters only polynomial in $n$. These 2-local Hamiltonia in particular are expressive enough to capture all quadratic unconstrained binary optimization (QUBO) problems.

The Variational Monte Carlo (VMC) \cite{mcmillan-pr65} is a heuristic algorithm which, given a local Hamiltonian $H$, produces an estimate for the minimal eigenvalue $\lambda_{\rm min}(H)$ and a description of an associated eigenvector.  By exploiting neural networks as trial wavefunctions, Carleo and Troyer \cite{carleo2017solving} showed that VMC can achieve state-of-the-art results for the ground state energies of physically important magnetic spin models. The domain of applicability of so-called neural-network quantum states has since been expanded considerably. It was recently shown, for example, that in the case of binary optimization, VMC is equivalent to Natural Evolution Strategies (NES) \cite{gomes2019classical} and state-of-the-art results can be achieved for the Max-Cut Hamiltonian \cite{zhao2020natural}, albeit at the expense of significantly increased computation time compared to the best known classical heuristic approximation algorithms. The slowdown is attributed to the gradient-based training loop within the VMC. In this step, the neural network parameters $\theta \in \mathbb{R}^d$ are guided in the direction of steepest descent of the following unbiased estimator of the Rayleigh quotient which upper bounds the lowest eigenvalue:
\begin{equation}\label{e:vmc}
    L(\theta) := \frac{\langle \psi_\theta | H | \psi_\theta \rangle}{\langle \psi_\theta | \psi_\theta \rangle}
    =
    \underset{x \sim |\psi_\theta(\cdot)|^2 }{\mathbb{E}}\left[\frac{\langle x| H | \psi_\theta\rangle}{\psi_\theta(x)}\right] \geq \lambda_{\min}(H) \enspace ,
\end{equation}
where $|x\rangle$ denotes the computational basis of $(\mathbb{C}^2)^{\otimes n}$ and the vector $|\psi_\theta\rangle \in (\mathbb{C}^2)^{\otimes n}$ denotes a neural-network quantum state in the sense that the amplitude $\psi_\theta(x) := \langle x | \psi_\theta \rangle$ is computed by a neural network with variational parameters $\theta$. The expectation value is taken with respect to the associated Born probabilities, which are proportional to $|\psi_\theta(x) |^2$. In practice, the objective function is optimized using a variant of stochastic minibatch gradient descent called Stochastic Reconfiguration \cite{sorella_aps98}.

In contrast to the canonical formulation of VMC, which accepts a single Hamiltonian as input, in this paper we propose the \emph{meta-VMC}, which asks for an approximation of the ground energy for an \emph{ensemble} of 2-local Hamiltonia $H_\tau$ indexed by a random disorder parameter $\tau$ sampled from a known distribution $\mathcal{T}$. The simplest strategy of retraining a separate neural-network quantum state from scratch for each realization of the disorder parameter $\tau$ is clearly impractical. The goal is thus shifted to finding a neural network that is maximally adaptive to new realizations of the disorder. For experimental support, we focus in this paper on the special case of QUBO problems, which can be identified with a Hamiltonian that is diagonal in the Pauli-$Z$ basis. These preliminary results are viewed as a stepping stone to the more physically interesting problem of random quantum spin models, in which we anticipate similar optimization considerations to apply.

The formulation of meta-VMC exhibits obvious parallels with \emph{meta-learning} or \emph{learning to learn} in the Machine Learning literature \cite{thrun2012learning}, where data from previously encountered learning tasks is employed to accelerate performance on new tasks, drawn from an underlying task distribution $\mathcal{T}$. In the language of meta-learning, $\tau$ indexes the learning task and $\mathcal{T}$ denotes the distribution over all tasks.  
In meta-VMC, we assume the task distribution is known to the learner; in contrast, conventional meta-learning assumes $\mathcal{T}$ is unknown but possesses sufficient regularity to render meta-learning feasible. 
A second, less significant distinction is that the per-task objective function for meta-VMC is an unbiased estimator for the population objective, whereas meta-learning typically focuses on Empirical Risk Minimization objectives, which suffer from nonzero bias.

\section{Theory}

A simple strategy that has proven successful in meta-learning of deep neural networks is multi-task transfer learning \cite{caruana1997multitask, baxter2000model}, which aims to learn an initialization for subsequent tasks by jointly optimizing the learning objective of multiple tasks simultaneously, using a minibatch training strategy that interleaves batches across the tasks. Multi-task learning is, however, prone to catastrophic interference \cite{mccloskey1989catastrophic}, making it unsuitable for generalization to the VMC. The problem is exemplified by some of the simplest examples of disordered spin systems: suppose $H_\tau$ is a random Hamiltonian whose expected value under the disorder parameter vanishes $\mathbb{E}_{\tau \sim \mathcal{T}} [H_\tau] = 0$. As a concrete example, consider the Sherington-Kirkpatrick Hamiltonian, in which $\tau$ represents a collection of i.i.d. centered Gaussian random variables $J_{ij} \sim N(0,1)$ representing the exchange energies. If we denote by $L_\tau$ the objective function corresponding to disorder parameter $\tau$, then the multi-task learning objective function, expressed in the population limit, is given by
\begin{equation}
    L_{\rm MTL}(\theta) := \underset{\tau \sim \mathcal{T}}{\mathbb{E}}\left[L_\tau(\theta)\right] 
    = \underset{\tau \sim \mathcal{T}}{\mathbb{E}} \left\{\underset{x \sim |\psi_\theta(\cdot)|^2 }{\mathbb{E}}\left[\frac{\langle x| H_\tau | \psi_\theta\rangle}{\psi_\theta(x)}\right]\right\} = \frac{\langle \psi_\theta | \mathbb{E}[H_\tau] | \psi_\theta\rangle}{\langle \psi_\theta |  \psi_\theta \rangle} = 0 \enspace .
\end{equation}
The fact that the multi-task learning objective loses dependence on $\theta$ in the population limit implies that the associated minibatch algorithm makes no progress asymptotically.

In order to define an objective function which is asymptotically non-vacuous and which promotes adaptation to new realizations of disorder, we propose to optimize the following meta-learning objective function, again presented in population form for simplicity \cite{andrychowicz2016learning},
\begin{equation}
    L_{\rm ML} (\theta) 
    := \underset{\tau \sim \mathcal{T}}{\mathbb{E}} \left[ L_\tau\big(U^t_\tau(\theta)\big) \right] 
    =
    \underset{\tau \sim \mathcal{T}}{\mathbb{E}} \big[ L_\tau\big(\underbrace{U_\tau \circ \cdots \circ U_\tau}_{\text{$t$ times}}(\theta)\big) \big]
     \enspace ,
\end{equation}
where $U^t_\tau : \mathbb{R}^d \to \mathbb{R}^d$ denotes the $t$-fold application of a task adaptation operator $U_\tau$, which in the simplest case of gradient descent with step size $\beta$, is given by $U_\tau(\theta) = \theta - \beta \nabla L_\tau(\theta)$. Optimization of the meta-learning objective $L_{\rm ML}$ ensures that when a new realization of the disorder parameter is drawn, the initialization performs well after performing one or more steps of gradient descent. Loosely speaking, meta-learning can be justified when one has a budget for running a few steps of gradient descent. 

The fact that the meta-learning objective function manages to avoid the catastrophic interference phenomenon can be illustrated by the following toy model\footnote{This quadratic model has also been analyzed in the context of convergence theory in \cite{fallah2020convergence}.}. Rather than considering the Rayleigh quotient \eqref{e:vmc}, consider the following ensemble of quadratic functions specified by a random positive-definite matrix $A \in \mathbb{R}^{d\times d}$ and a random vector $b \in \mathbb{R}^d$,
\begin{equation}
    L_\tau(\theta) = \frac{1}{2} \langle \theta, A  \, \theta \rangle - \langle b, \theta \rangle \enspace ,
\end{equation}
where the random variable $\tau = (A, b)$ now corresponds to the task label.
In the simplest setting of single-step ($t=1$) meta-learning with vanilla update operator $U_\tau (\theta) = \theta - \beta \nabla L_\tau(\theta)$, the optimal solution of the multi-task and meta-learning objectives can be found in closed form,
\begin{align}
    \argmin_{\theta \in \mathbb{R}^d} L_{\rm ML}(\theta) & = \mathbb{E} \big[A(I-\beta A)^2\big]^{-1} \mathbb{E} \big[(I - \beta A)^2 b\big] \enspace . 
\end{align}
In the limit $\beta \to 0$ corresponding to multi-task learning, the optimal solution is found to only depend on the mean value of the random variable $\tau$, whereas the meta-learner ($\beta > 0$) exploits information in the higher-order moments of $\tau$.

In the case of meta-VMC, we consider gradient-based optimization. Specifically, we focus on model-agnostic meta-learning (MAML) \cite{finn2017model} which is a gradient-based algorithm that has been proposed for optimizing the meta-learning objective. Straightforward application of the chain rule gives rise to the following gradient estimator for the meta-learning objective,
\begin{equation}
    \nabla L_{\rm ML}(\theta) = \mathbb{E}_{\tau \sim \mathcal{T}}\big[(U_\tau^t)'(\theta) \nabla L_\tau\big(U_\tau^t(\theta)\big)\big] \enspace ,
\end{equation}
where $(U_\tau^t)'(\theta)$ denotes the Jacobian matrix of the function $U_\tau^t : \mathbb{R}^d \to \mathbb{R}^d$.
The pseudocode for MAML is outlined in Algorithm \ref{algo:pseudocode}. In order to facilitate readability, we have presented the algorithm with batching only in the task index, leaving the remaining expectation values (with respect to Born probabilities) in population form. In a practical algorithm, the intermediate variables $\theta_\tau$ and $\nabla_\tau$ are estimated stochastically using independent batches of data generated by the same task $\tau$. Since the computation of the Jacobian involves an expensive back-propagation, first-order MAML (foMAML) has been proposed (e.g., \cite{finn2017model, nichol2018first}), which is a simplification of MAML in which the Jacobian matrix is approximated by the identity matrix.
\begin{algorithm}[H]
\textbf{Input:} Matrix ensemble $\mathcal{T}$, adaptation operator $U_\tau$, adaptation steps $t$\\
Initialize $\theta$ \\
\While{not done}{
    Sample batch of disorder parameters $B \overset{\rm iid}{\sim} \mathcal{T}$ \\
    \For{each disorder parameter $\tau \in B$}{
    $\theta_\tau = U^t_\tau (\theta)$ \\
    $\nabla_\tau = (U_\tau^t)'(\theta) \nabla L_\tau(\theta_\tau)$
    }
    $\nabla = \frac{1}{|B|} \sum_{\tau \in B} \nabla_\tau$ \\
    $\theta \gets \textsc{Optimizer}(\theta,\nabla)$
}
\caption{MAML \cite{finn2017model}  adapted to meta-VMC (batched over tasks).\label{algo:pseudocode}}
\end{algorithm}
\section{Relationship with previous work}
In this section, we differentiate our proposal from the uses of meta-learning that have been proposed in the quantum computing literature. In \cite{wilson2019optimizing}, for example, meta-learning has been proposed to mitigate various sources of noise, specifically shot noise and parameter noise. In the context of VMC, shot noise is analogous to variance associated with finite minibatches, whereas parameter noise has no clear analogue. Ref.~\cite{verdon2019learning} is the most similar to ours in that they consider meta-learning from known distributions. They differ by the choice to focus on variational quantum algorithms such as VQE and QAOA and by the fact that they do not use model-agnostic meta-learning. Instead, the meta-learning outer-loop involves training a separate recurrent neural network, similar to \cite{andrychowicz2016learning}.

\section{Experiments}
The experiments focus on the Max-Cut problem which is defined in terms of a simple, undirected graph $G=(V,E)$ with binary-indicator adjacency matrix $J$. In order to attack the Max-Cut problem with the VMC, we encode the solution in the ground state of the following classical antiferromagnetic Ising Hamiltonian with exchange interaction energy matrix given by $J$,
\begin{equation}
H_\tau = \sum_{1\leq i < j\leq n} J_{ij} Z_i Z_j \enspace ,
\end{equation}
where $Z_i$ denotes the Pauli-$Z$ operator acting locally to the $i$th qubit. Since the Max-Cut Hamiltonian is diagonal in the Pauli-$Z$ basis, it acts as a multiplication operator and the ground state can be chosen as a computational basis vector $|x \rangle $ corresponding to a maximal cut. The variational wavefunction was chosen to be a real-valued Boltzmann machine with $n$ hidden units. The task distribution $\mathcal{T}$ defining the ensemble of Max-Cut Hamiltonians was defined by the following procedure. The adjacency matrix $A$ for a Bernoulli random graph with edge probability $0.5$ was first chosen and fixed throughout the experiments. Sampling an adjacency matrix from the task distribution $\mathcal{T}$ is performed by rounding the matrix $A + (X + X^T)/2$ to a 0/1 matrix, where $X$ denotes an $n \times n$ matrix with entrywise Gaussian noise $X_{ij} \sim N(0,\sigma^2)$.
The hyperparameter controlling the task diversity for this ensemble is thus the variance $\sigma^2$. 

\begin{figure}[htbp!]
\small
\centering
\vspace{-0.5cm}
\includegraphics[width=1.0\linewidth]{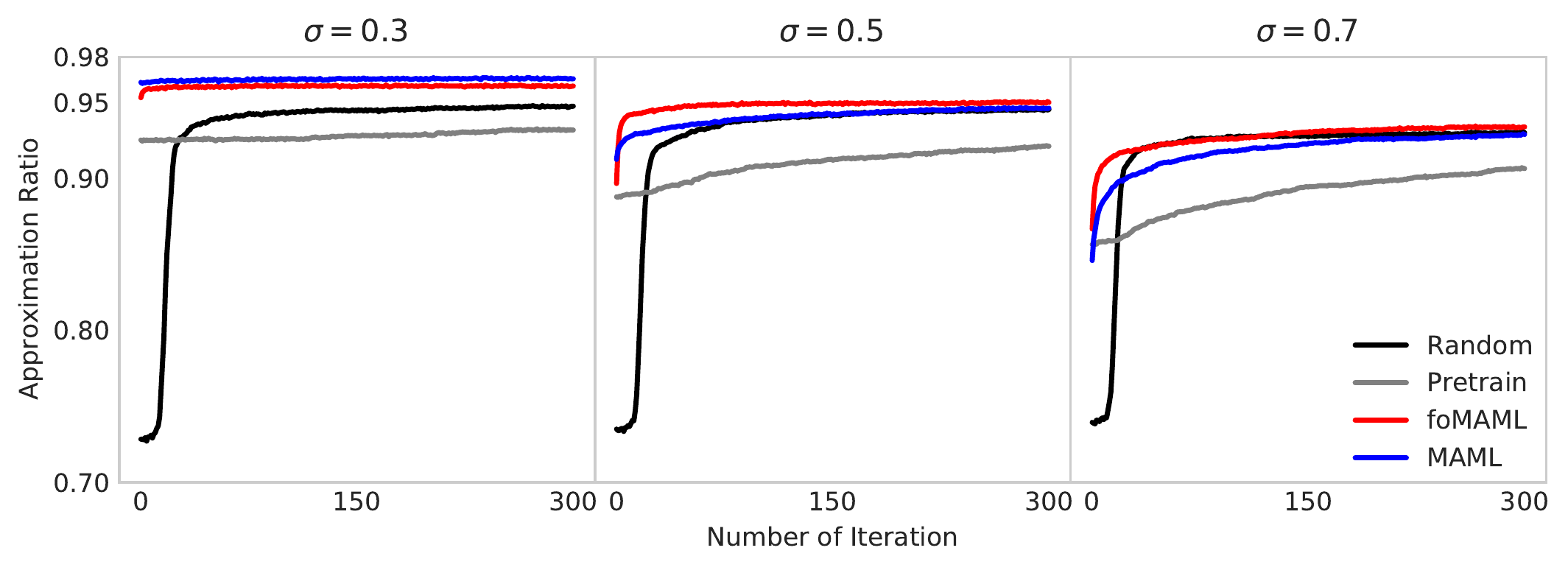}
\caption{VMC training curves for Max-Cut instances ($|V|=50$) using different parameter initialization strategies. Two baselines are considered: \textbf{Random} refers to training from scratch and \textbf{Pretrain} initializes with the model pre-trained with the base graph. Experiments were conducted using three task distributions characterized by $\sigma\in \{0.3,0.5,0.7\}$. For each task distribution, $N_{\rm test} = 32$ graph instances are sampled for testing. We train the models on all testing graphs for 300 VMC iterations and evaluate the performance with the averaged approximation ratio, which is defined to be the ratio between the cut number and optimal value of the SDP relaxation. Models initialized from MAML and foMAML can discover near-optimal solutions within very few iterations, whereas pre-trained models get stuck in the local maximum. Moreover, MAML and foMAML provide better model parameter initializations that outperform the baselines in the long run. The advantage diminishes, however, as $\sigma$ gets large.}
\label{fig:meta_vmc}
\vspace{-0.5cm}
\end{figure}

During training, each iteration of the meta-learning loop involved independently sampling a batch of $16$ tasks from $\mathcal{T}$. During testing, $N_{\rm test} = 32$ adjacency matrices were sampled from $\mathcal{T}$ and fixed for evaluation purposes.
The inner loop used $t=15$ iterations of vanilla SGD (step size $\beta =0.01$ batch size 128), while the outer loop training used 100 iterations of vanilla SGD (step size $\alpha =0.01$, batch size 16).
The meta-learning experiments were conducted using MAML \cite{finn2017model} and first-order MAML (foMAML) \cite{finn2017model, nichol2018first}. For baselines, we compared against training a neural-network quantum state from scratch as well as a pre-trained initialization with fine-tuning. The learning curves, illustrated for different values of $\sigma$ in Fig.~\ref{fig:meta_vmc}, clearly show that model-agnostic meta-learning dramatically accelerates training compared to the baselines on the $N_{\rm test}$ testing Max-Cut instances, consistent with the goal of MAML in promoting adaptivity. The networks trained using MAML also found larger cut values compared to the baselines which appear to converge prematurely to suboptimal states.

\section{Discussion and future directions}
The preliminary experimental results for Max-Cut ensembles indicate that MAML effectively solves meta-VMC by accelerating training and improving convergence.  While the Max-Cut problem is exactly solvable for the graph sizes considered here (e.g., by the Branch and Bound method \cite{rendl2010solving}), our work paves the way to investigate physically-interesting matrix ensembles that cannot be diagonalized by a local change of basis. The ideas presented in this paper naturally extend also to variational quantum algorithms (VQAs) such as the variational quantum eigensolver. The key difference in the case of VQAs is that the denominator in the Rayleigh quotient \eqref{e:vmc} is automatically normalized $\langle \psi_\theta | \psi_\theta \rangle = 1$ and stochastic estimation of the quantum expectation value $\langle \psi_\theta | H_\tau | \psi_\theta \rangle$ involves performing measurements in multiple bases if the Hamiltonian contains non-commuting terms. The exploration of meta-VQA and associated learning algorithms is left to future work.

\section*{Broader Impact}
The authors have not identified any ethical impacts or future societal consequences of this work.
% Authors are required to include a statement of the broader impact of their work, including its ethical aspects and future societal consequences. 
% Authors should discuss both positive and negative outcomes, if any. For instance, authors should discuss a) 
% who may benefit from this research, b) who may be put at disadvantage from this research, c) what are the consequences of failure of the system, and d) whether the task/method leverages
% biases in the data. If authors believe this is not applicable to them, authors can simply state this.

% Use unnumbered first level headings for this section, which should go at the end of the paper. {\bf Note that this section does not count towards the eight pages of content that are allowed.}

\begin{ack}
The authors would like to thank Giuseppe Carleo for many helpful discussions. Authors gratefully acknowledge support from NSF under grant DMS-2038030.
\end{ack}

% Use unnumbered first level headings for the acknowledgments. All acknowledgments
% More information about this disclosure can be found at: \url{https://neurips.cc/Conferences/2020/PaperInformation/FundingDisclosure}.

% Do {\bf not} include this section in the anonymized submission, only in the final paper. You can use the \texttt{ack} environment provided in the style file to autmoatically hide this section in the anonymized submission.
% \end{ack}

\bibliographystyle{plain}
\bibliography{references}

\end{document}